\documentclass[aps,prb,twocolumn,floatfix]{revtex4-1}
\usepackage{graphicx,epsfig,array,color}
\usepackage{color,graphicx,amssymb,amsmath}
\usepackage[utf8]{inputenc}
\usepackage[caption=false]{subfig}
\bibpunct{[}{]}{,}{n}{}{}
\begin{document}
\title{Magnetic properties and pairing tendencies of the \\ 
iron-based superconducting ladder BaFe$_2$S$_3$: \\
combined \emph{ab initio} and density matrix renormalization group study }
\author{Niravkumar D. Patel$^{1,2}$}
\author{Alberto Nocera$^3$}
\author{Gonzalo Alvarez$^3$}
\author{Ryotaro Arita$^{4,5}$}
\author{Adriana Moreo$^{1,2}$}
\author{Elbio Dagotto$^{1,2}$}

\affiliation{$^1$Department of Physics and Astronomy, The University of Tennessee, Knoxville, Tennessee 37996, USA}
\affiliation{$^2$Materials Science and Technology Division, Oak Ridge National Laboratory, Oak Ridge, Tennessee 37831, USA}
\affiliation{$^3$Computer Science $\&$ Mathematics Division and Center for Nanophase Materials Sciences, Oak Ridge National Laboratory, Oak Ridge,Tennessee 37831, USA}
\affiliation{$^4$JST, ERATO, Isobe Degenerate $\pi$-Integration Project, Aoba-ku, Sendai 980-8577, Japan}
\affiliation{$^5$RIKEN Center for Emergent Matter Science, Wako, Saitama 351-098, Japan}

\begin{abstract}

The recent discovery of superconductivity under high pressure in the two-leg ladder compound 
BaFe$_2$S$_3$ [H. Takahashi {\it et al.}, Nature Materials {\bf 14}, 1008 (2015)] opens a broad avenue of
research, because it represents the first report of pairing tendencies
in a quasi one-dimensional iron-based high critical temperature superconductor.
Similarly as in the case of the cuprates, ladders and chains can be far more
accurately studied using many-body techniques and model Hamiltonians than their
layered counterparts, particularly if several orbitals are active. In this publication,
we derive a two-orbital Hubbard model from first principles that describes individual
ladders of BaFe$_2$S$_3$. The model is studied with the density matrix renormalization group.
These first reported results are exciting for two reasons: 
{\it (i)} at half-filling, ferromagnetic order emerges as the
dominant magnetic pattern along the rungs of the ladder, 
and antiferromagnetic order along the legs, in excellent agreement with
neutron experiments; {\it (ii)} with hole doping,
pairs form in the strong coupling regime, as found by
studying the binding energy of two holes doped on the half-filled system. In addition,
orbital selective Mott phase characteristics develop with doping, with only one Wannier
orbital receiving the hole carriers while the other remains half-filled.
These results suggest that the analysis of models for iron-based two-leg ladders 
could clarify the origin of pairing tendencies and other exotic properties
of iron-based high critical temperature superconductors in general.
\end{abstract}

\maketitle

\section{Introduction} 

The understanding of the high critical temperature ($T_c$)
superconductors based on iron continues attracting 
the attention of the condensed matter community~\cite{DCJohnston,Stewart,Peter,Scalapino,Chubukov,Dainature,RMP}.
It is widely believed that these studies may not only have
potential technological applications, but they may also
shed light on other high-$T_c$ superconductors such as those
based on copper. While early theoretical studies of
the iron-based compounds were guided by simple Fermi
surface nesting ideas that may have captured important
properties of these materials such as the symmetry of
the superconducting state, more recent investigations are increasingly
suggesting that the effect of Coulombic repulsion between
electrons cannot be neglected~\cite{Dainature}. For example, there are 
compounds that are superconducting but do not have hole
pockets, and thus no nesting effects, 
at the Fermi surface~\cite{nFSn}. There are also materials with robust
magnetic local moments even at room temperature~\cite{moments1,moments2}, in disagreement
with weak coupling perspectives where the formation of moments 
and their long-range order occur simultaneously upon cooling. Moreover,
complex spin arrangements have been unveiled in several
materials, as recently reviewed~\cite{RMP}. All these results 
suggest that repulsive interactions
between electrons are important to fully understand these
compounds' properties. However, the theoretical analysis of
multiorbital Hubbard models is challenging because of
the absence of reliable many-body tools to study their properties
in layered systems.
In particular, thus far the only theoretical evidence that superconductivity
can be induced in these compounds via antiferromagnetic (AFM) fluctuations 
relies exclusively on BCS
gap equations and random phase approximation techniques. Can
we generate more robust theoretical evidence for AFM-based superconductivity
in these compounds?

In the context of the copper-oxide high-$T_c$ 
superconductors, finding crystal structures simpler than layers 
but still with intriguing quantum mechanical many-body properties proved to be
a fruitful path for progress in that field. One of the reasons is that
theorists can perform model Hamiltonian calculations with more accuracy
in e.g. quasi one-dimensional systems. In fact, spin-1/2 Cu-oxide two-leg ladders
have been much studied  in cuprates because  of  their  unusual  spin
gap,  induced  by  the  ladder  geometry~\cite{ladder1,ladder2,ladder3}.  The  Cu-oxide-ladder
spin state is dominated by rung spin-singlets, and it was theoretically 
predicted that such a system should have a tendency
to superconductivity upon doping. This was verified in high pressure experiments
at $\sim$3~GPa for the case of Sr$_{0.4}$Ca$_{13.6}$Cu$_{24}$O$_{41.84}$,
reporting a critical temperature of 12~K~\cite{uehara}. 
Due to its quasi one-dimensional character,
it was possible to employ a variety of accurate many-body techniques for ladders, showing agreement
between theoretical predictions and experimental results, 
an agreement that has provided considerable support to the notion that superconductivity in cuprates originates in AFM fluctuations.

These important earlier results in the context of copper-oxide ladders 
suggest that progress in the understanding of iron-based superconductors
would be possible if similar quasi one-dimensional structures could be
prepared and theoretically studied. For  this  reason  considerable
interest was generated by recent studies of BaFe$_2$Se$_3$
because this material contains double chains made of
[Fe$_2$Se$_3$]$^{2-}$ blocks separated 
by Ba~\cite{Feladder1,Feladder2,Feladder3,Feladder4,Feladder5,Feladder6,Feladder7,Feladder8,Feladder9}. 
The resulting structure contains two extended Fe-Fe directions
(the ``legs'') connected by Fe-Fe bonds of similar strength (the ``rungs'')
thus defining two-leg ladders very similar to those in the cuprates. A difference
is that the Cu-Cu bridge in the cuprates' ladders 
is made by an oxygen in between the coppers, while in chalcogenides
the bridges between irons are provided by selenium which is located up and down the middle
of the iron plaquettes. Thus, as in their two dimensional counterparts, electronic
hoppings of similar strength are to be expected for the chalcogenides
not only along legs and rungs, but also along the plaquette diagonals.

BaFe$_2$Se$_3$ is  an  insulator,  with  an activation energy 
between 0.13~eV~\cite{Feladder4} and 0.178~eV~\cite{Feladder2}, 
long-range AFM order at
$\sim$250~K induced by weak residual interladder interactions,
and low-temperature magnetic 
moments $\sim$2.8~$\mu_B$~\cite{Feladder1,Feladder2,Feladder3}. Remarkably,
neutron diffraction studies reported a dominant magnetic
order  at  low temperature involving  blocks  of  four  iron  atoms  with
their  moments  aligned,  coupled  antiferromagnetically  along the ladder
direction~\cite{Feladder1,Feladder4}. 
When K replaces Ba, thus leading to KFe$_2$Se$_3$, the
magnetic state changes to an arrangement where the  spins  in  the  same  rung  are
coupled ferromagnetically but they are antiferromagnetically
ordered in the long ladder direction~\cite{Feladder5}. Theoretical studies primarily 
employing the Hartree Fock approximation~\cite{Feladder7} unveiled a rich phase diagram
for two-leg ladder multiorbital Hubbard models, with a plethora of phases
including both states already found in the Ba- and K-based ladders as well
as several other competitors. These exotic spin arrangements arise 
from frustrating tendencies between the staggered AFM state that dominates
at small Hund coupling and the ferromagnetic (FM) state stable at large
Hund coupling~\cite{Feladder7}. Hartree-Fock results for layers~\cite{luo2D} 
and chains~\cite{luo1D}  also suggest a complex landscape of competing magnetic states 
in those geometries.

Recently, an unexpected experimental result has been reported using BaFe$_2$S$_3$~\cite{NatMatSC,ohgushi},
where S replaces Se but keeps the two-leg ladder structure the same. This material was found
to become superconducting at a pressure above 10 GPa with an optimal critical temperature
$T_c = 24$~K. The parent compound, i.e. the same material but at ambient pressure,
is a Mott insulator with the same magnetic order as KFe$_2$Se$_3$ namely involving
FM rung and AFM leg spin correlations with a critical temperature $\sim 120$~K,
according to power neutron diffraction studies~\cite{NatMatSC}. These
discoveries unveiled the first iron-based superconductor that does not rely on a
square lattice structure of irons, 
opening an intriguing avenue of research similar to
the one opened
with the discovery of superconductivity in Cu oxide ladders in the context of the cuprates.

The present publication introduces a two-orbital Hubbard model for a two-leg ladder of BaFe$_2$S$_3$, based on {\it ab initio} calculations. This model is subsequently solved 
computationally using the density matrix renormalization group (DMRG) technique~\cite{SRWhite}.
Our main results are two folded. First, we show that at half-filling with two electrons
per iron, and using clusters as large as 16$\times$2,  
there is a robust evidence for the same magnetic order found experimentally
involving FM rungs and AFM leg correlations. This magnetic state becomes robust at
intermediate and strong Hubbard couplings, in agreement with the growing perception that
these materials are not in the weak coupling regime. Second, we assume that in the 
experiments~\cite{NatMatSC} the high pressure alters the band structure in such a manner that
the individual ladders become hole doped, although the insulator-superconductor
transition could also be bandwidth-controlled~\cite{NatMatSC}. In the Cu-oxide based
ladders studied some years ago, experiments showed~\cite{piskunov} 
that indeed pressure alters the amount
of mobile electrons residing in the two-leg ladders in such a manner that the 
superconducting state is reached effectively by hole doping of the ladders. Here
we simply assume that a similar physics 
occurs in the iron-based ladders and focus on their
hole doping.  In fact,
studying the cases of one, two, and four holes
we have found pairing tendencies when using an 8$\times$2 cluster in
the strong coupling regime $U/W \simeq 2$, $W$ being the tight binding electronic bandwidth. 
The complexity of the Hamiltonian with
two active orbitals and a tight-binding term that must include plaquette diagonals hoppings
renders the DMRG calculation so computing time demanding 
that a confirmation of the pairing tendency beyond 8$\times$2 is
not possible at present with the DMRG technique 
and our available computer resources. Nevertheless, the pairing indications
we have observed are promising and suggestive that the theoretical study of
iron-based two-leg ladders may illuminate the understanding of iron-based superconductors
using many-body techniques beyond the diagrammatic random phase approximation. 

The organization of the manuscript is as follows. Section~II 
provides details of the
{\it ab initio} calculations. Section III contains the actual model
used, many-body technique, and observables studied. Section IV presents our main
results, organized separately for zero, one, and two holes, the latter including
binding energies. Finally, Sec.~V contains our main conclusions.

\section{\emph{Ab Initio} Calculations}

This section presents the details of the derivation
of
the multiorbital Hubbard model for the BaFe$_2$S$_3$ ladder from first principles, to be used later in Sections III and IV. 
Following the procedure described in Ref.~\cite{aritaPRB15}, first a 
calculation is performed based on the generalized gradient approximation
with the {\sc Quantum Espresso} package~\cite{QE}. 
There we employed the exchange-correlation functional proposed by 
Perdew, Burke and Ernzerhof~\cite{PBE}, a plane-wave basis set
with a cutoff energy of 40 Ry, and an 8$\times$8$\times$8 {\bf k}-mesh 
for the first Brillouin zone (BZ). 
As for the lattice constants, we used the experimental values
$a$ = 8.78 \AA, $b$ = 11.23 \AA, $c$ = 5.29 \AA\ for the ambient
pressure case and reduced them by 4.0\%,
8.0\%, and 3.4\%, respectively, for pressure $P$=12.4 GPa~\cite{HirataPRB15}.
The space group of the system is
$Cmcm$, and the atomic positions of Ba(4c), Fe(8e), S(4c), and
S(8g) are (0.0, 0.686, 0.25), (0.154, 0.0, 0.0), (0.0, 0.116, 0.25),
and (0.208, 0.378, 0.25), respectively~\cite{HirataPRB15}.
Because the magnetic properties will be considered when we 
solve the effective two-orbital Hubbard model in the following sections,
magnetism was not included in the derivation of the model from first principles~\cite{SuzukiPRB15}.

\begin{figure}[!ht]
\centering{
\includegraphics[width=8.5cm, height=5.5cm, clip=true]{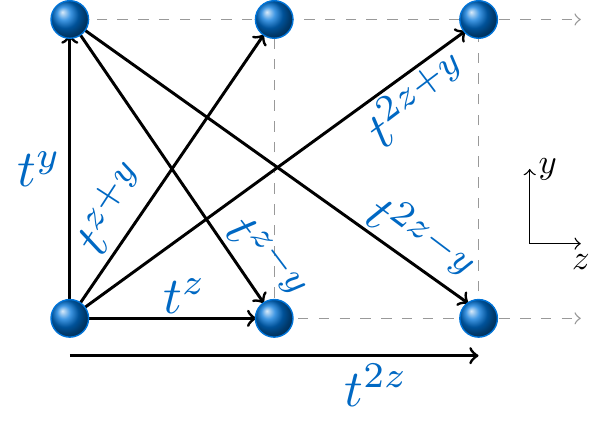}}
\vspace{-0.3cm}
\caption{(color online) Schematic representation of the directions of electronic hopping 
for the two-leg ladder model considered here. 
The legs of the ladder are arranged in the $z$-direction 
while the rungs are in the $y$-direction. The hoppings to next-nearest neighbor rungs 
(i.e. $t^{2z},t^{2z-y},t^{2z+y}$) are dubbed 
the ``long-range hoppings''. In total, there are seven different hopping 
directions shown.
}
\label{Fig1.FeSLattice}
\end{figure}

After this initial setup, 
we constructed two Wannier functions for each Fe atom in the unit cell using the
{\sc Wannier90} package~\cite{wannier90}.
One resulting Wannier orbital mainly consists of the standard $d_{x^2-y^2}$ orbital (orbital $a$ below)
while the other one is primarily made from the standard
$d_{xz}$ orbital (orbital $b$ below). The two Wannier orbitals employed here do not have a high symmetry, because the
$d_{x^2-y^2}$ or $d_{xz}$ orbitals significantly 
hybridize with other $d$-orbitals or S-$p$ orbitals. 
In particular, for the latter Wannier 
orbital there is a substantial contribution of the
canonical $d_{xy}$ orbital.
To construct explicitly 
these orbitals, we have to disentangle their complicated electronic structure.
In order to preserve accurately the properties of the low-energy band dispersion, 
we introduced a ``frozen'' energy window~\cite{Vanderbilt} as large as 
(-0.3 eV, 0.2 eV) with respect to $E_F$, on top of the ordinary ``global'' energy window
(-1.2 eV, 1.5 eV).

After we constructed an 8 band model, we further simplified this model by unfolding the
BZ along the $k_z$ direction. Namely, by introducing a local gauge transformation 
for one of the two Wannier orbitals to change its sign, we can expand the band
dispersion from $\Gamma$ to $Z$ in the original BZ, and construct a 4 band model~\cite{aritaPRB15}.

Finally, we want to derive an effective ladder model from the
4-band model, namely we wish to arrive to a model restricted to a two-leg ladder. 
One possibility is to neglect all inter-ladder electron hopping transfers and
just focus on the intra-ladder transfers. However, the bandwidth in the $k_x$-$k_y$ plane is
not necessarily small (as large as 500 meV at maximum). Thus, in the present study
the effect of inter-ladder transfers is taken into account 
by considering their average, i.e.,
we construct our ladder model from the 4-band Hamiltonian 
by considering the case $k_x$=$k_y$=0.

\section{Model and Method}

This section explicitly provides the multiorbital Hubbard model derived by the procedure
explained before, while Section IV will present 
the magnetic properties and pairing tendencies of BaFe$_2$S$_3$. 
The model studied here breaks up into
kinetic energy and interaction terms:
$H=H_{K}+H_{int}$. The tight-binding kinetic energy portion is

\begin{equation} \label{Eq1:TightBinding}
 H_{k}= \sum_{\substack{ i \sigma \\ \gamma \gamma' \alpha }} 
 		t^{\vec{\alpha}}_{\gamma \gamma'} 
 		(c_{i\sigma\gamma}^{\dagger} c_{i+\vec{\alpha}\sigma\gamma'} 
 		+ H.c. ) 
 		+ \sum_{ i \gamma \sigma} \Delta_{\gamma} n_{i \gamma \sigma},
\end{equation}
where the first term represents the  hopping of an electron from 
site $i$ of a two-leg ladder and orbital $\gamma$ to site
$i + \vec{\alpha}$ and orbital $\gamma'$. The vector $\vec{\alpha}$ indicates
the many different directions possible for the electronic hopping, 
as shown in the ladder sketch Fig.~\ref{Fig1.FeSLattice}. 
We use a two orbital model
where we label the down-folded~\cite{aritaPRB15} orbitals as $a$ and $b$ 
(i.e. $\gamma$ and $\gamma'$ are restricted to $a$ and $b$). 
$\Delta_{\gamma}$ represents the 
crystal-field splitting of orbital $\gamma$. 
There are two sets of 
hopping parameters obtained from fitting the \textit{ab initio} 
down-folded band structure calculations at different pressures.
The crystal fields at $P=0.0$ GPa are $\Delta_{a} =  0.308$
and $\Delta_{b} = -0.229$ (eV units used from now on) while the associated
hopping amplitudes are

\begin{equation} \label{Eq2:P0}
\hspace*{-1cm} 
\begin{split}
t^{z} &= \begin{bmatrix}
    -0.215 & -0.149\\
    +0.149 & +0.153
    \end{bmatrix}, \\
    \\
t^{y} &= \begin{bmatrix}
    -0.012 & 0.000\\
     0.000 & +0.153
    \end{bmatrix},\\
    \\
t^{z+y} = t^{z-y} &= \begin{bmatrix}
    +0.075 & +0.174\\
    -0.174 & +0.083
    \end{bmatrix}, \\
    \\
t^{2z} &= \begin{bmatrix}
    -0.137 & +0.004\\
    -0.004 & +0.037
    \end{bmatrix},\\
    \\
t^{2z+y} = t^{2z-y} &= \begin{bmatrix}
    -0.007 & +0.016\\
    -0.016 & -0.041
    \end{bmatrix}, \\
\end{split}
\end{equation}
\noindent while the crystal fields at $P = 12.36$ GPa are
$\Delta_{a} =  0.423$ and $\Delta_{b} = -0.314$, 
with associated hoppings amplitudes

\begin{equation} \label{Eq3:P12}
\hspace*{-1cm} 
\begin{split}
t^{z} &= \begin{bmatrix}
    -0.334 & -0.177\\
    +0.177 & +0.212
    \end{bmatrix}, \\
    \\
t^{y} &= \begin{bmatrix}
    -0.024 & 0.000\\
     0.000 & +0.216
    \end{bmatrix}\\
    \\
t^{z+y} = t^{z-y} &= \begin{bmatrix}
    +0.085 & +0.216\\
    -0.216 & +0.109
    \end{bmatrix}, \\
    \\
t^{2z} &= \begin{bmatrix}
    -0.171 & -0.011\\
    +0.011 & +0.035
    \end{bmatrix},\\
    \\
t^{2z+y} = t^{2z-y} &= \begin{bmatrix}
     0.000 & +0.042\\
    -0.042 & -0.044
    \end{bmatrix}. \\
\end{split}
\end{equation}

Figure~2~(a,b) show
the single particle spectrum, calculated to illustrate the band structure 
at both $P = 0.0$ and $P = 12.36$ GPa using all the hoppings in Eqs.~\ref{Eq2:P0} and \ref{Eq3:P12} (``long range hoppings''). 
In Fig.~3~(a,b) similar results are
presented but using only hoppings up to nearest neighbor rungs (``short range hoppings'').
The band structures in both cases are similar. However, some discrepancies occur.
For instance at the edges, such as $k_z$=0 and $\pi$, the short range hoppings present
degeneracies (or near degeneracies) 
that are split in the long range case. It is unclear if these 
details are significant or not, and without performing the DMRG calculations in both
cases explicilty this issue cannot be answered conclusively. Here we simply wish to alert the readers of
these small differences for completeness.

\begin{figure}[!ht]
\centering{
\includegraphics[width=8.0cm, height=8.0cm, clip=true, 
		trim={0.05cm 0.0cm 0.0cm 0.0cm}]
		{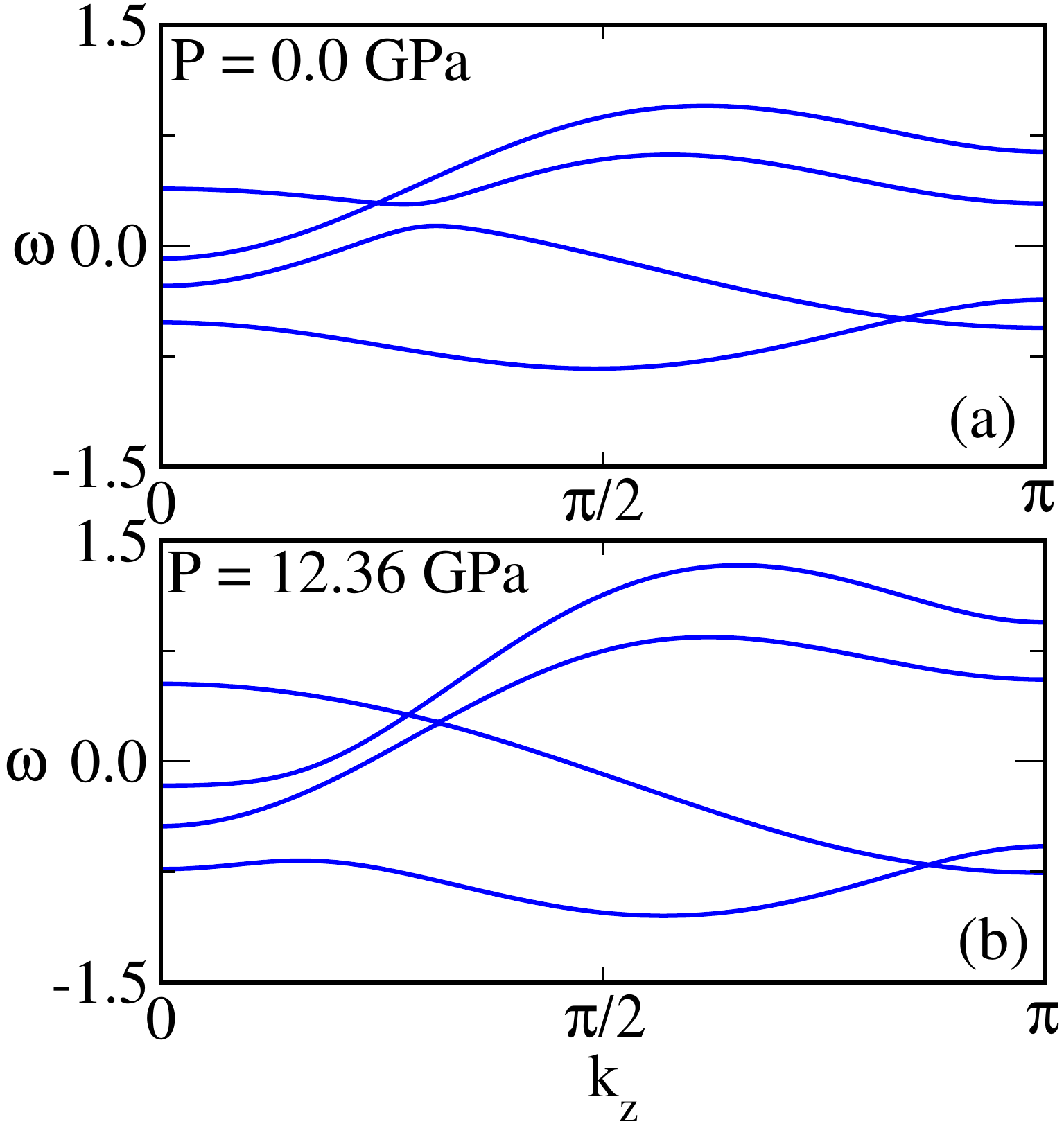}}
\caption{(color online) Tight-binding band structure ($U/W=0.0$)
involving hoppings up to the next-nearest neighbor 
rungs (i.e. long-range hoppings) for pressures of 
(a) $0.0$ and (b) $12.36$ GPa.
The chemical potential is at zero energy for
half-filling.
}
\label{Fig2.LongRange_Band}
\end{figure}

\begin{figure}[!ht]
\centering{
\includegraphics[width=8.0cm, height=8.0cm, clip=true, 
				trim={0.05cm 0.0cm 0.0cm 0.0cm}]			
				{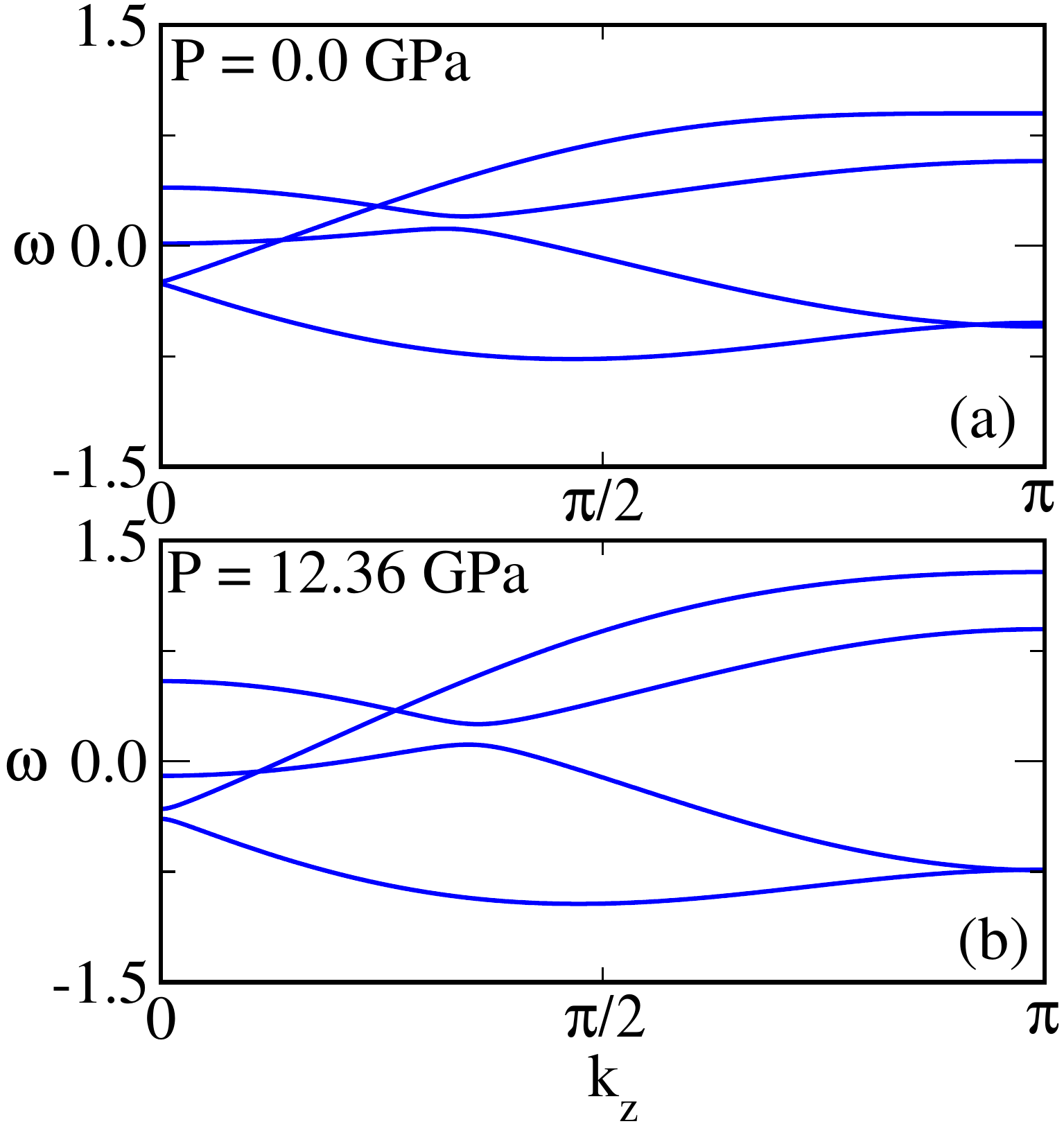}
			}
\caption{(color online) Tight-binding band structure ($U/W=0.0$)
involving hoppings only up to the nearest neighbor 
rungs (i.e. short-range hoppings) for pressures of (a) $0.0$ 
and (b) $12.36$ GPa.
The chemical potential is at zero energy for
half-filling.
}
\label{Fig3.ShortRange_Band}
\end{figure}
 
The electronic interaction portion of the Hamiltonian
\begin{equation}
 \begin{split}
 H_{int}& =U\sum_{i\gamma} n_{i\uparrow\gamma} n_{i\downarrow\gamma} + (U'- \frac{J}{2}) \sum_{\substack{i \\
 \gamma < \gamma'}} n_{i\gamma} n_{i\gamma'} \\ 
 &- 2J \sum_{\substack{i \\
 \gamma < \gamma'}} \textbf{S}_{i\gamma}.\textbf{S}_{i\gamma'} + 
 J \sum_{\substack{i \\
 \gamma < \gamma'}} (P_{i\gamma}^{\dagger} P_{i\gamma'} + \text{h.c.})
 \end{split}
\end{equation}
contains the standard intra-orbital 
Hubbard repulsion $U$ and the Hund's coupling $J$. The operator $\textbf{S}_{i\gamma}$ ($n_{i\gamma}$) is the total spin (electronic density) for orbital $\gamma$ at site $i$. $P_{i\gamma}^{\dagger}$ ($P_{i\gamma}$) are the pair creation (annihilation) operators. The standard 
relation $U' = U - 2J$ is assumed. The operators are defined in terms of the creation and annihilation fermion operators as 
\begin{equation}
\textbf{S}_{i\gamma} = \sum_{\sigma\sigma'} c_{i\sigma\gamma}^{\dagger} \sigma_{\sigma\sigma'} c_{i\sigma'\gamma},
\end{equation}
$n_{i\sigma\gamma} = c_{i\sigma\gamma}^{\dagger} c_{i\sigma\gamma} $, and $P_{i\gamma}=c_{i\downarrow\gamma} c_{i\uparrow\gamma}$. The half-filling electronic density corresponds to two electrons 
per site.

We use the ground state density matrix renormalization group (DMRG) technique 
with open boundary conditions 
in order to study the BaFe$_2$S$_3$ ladder using the 
two-orbital Hubbard model previously defined. 
DMRG grows the lattice by adding  
sites in a ``snake-like'' geometry. 
We have studied in detail a ladder size of $8\times2$ 
with up to four holes doped over a half-filled system. Calculations involving 
$12\times2$ and $16\times2$ ladders are also presented at half-filling. 
Keeping up to $800$ states, the typical value of the discarded weight (truncation error) 
is of the order of $10^{-5}$ for the doping cases studied.
Within this level of error observables are converged. For 
the two holes case we can reach a similar accuracy only for 8$\times$2 lattices, 
and for this reason our study of binding energies
is restricted to those lattices.
With typical computer resources, 
$8\times2$ ladder simulations with $m = 800$ states require 3-4 days at half-filling.
The two holes doped case needs more than a week, even if using only short-range hoppings.
The use of the long-range hoppings
substantially increases the time required for convergence. 
This is because of three different reasons.
First, the most time consuming part of the DMRG process is 
computing the Hamiltonian connections, and with long-range hoppings one has to sum a
large number of terms in the Hamiltonian.
Second, at fixed on-site interactions ($U$ and $J$) and fixed density, 
the difficulty of the DMRG scales exponentially with the
number of connections between system and environment,
when a lattice is split in the middle. 
Third, our on-site Hilbert space is large due to the presence of two
orbitals. 
In fact, the system we have studied can be translated into a one-dimensional 
one-orbital Hubbard model with hoppings up to 12$^{th}$ neighbors.
This illustrates the substantial numerical effort presented here at the limit
of what can be done with modern many-body computational techniques. We also want
to remark that perhaps more modern versions of DMRG, such as those involving matrix
product operators, may alleviate the effort needed in the present problem. In fact,
recently $S=1/2$ ladders including dipolar interactions were
studied with up to 400 rungs with this method~\cite{MPO}. 

We will present a variety of charge and magnetic observables for doping of up to  four 
holes on the half-filled system. The average occupation number of each orbital is
\begin{equation}
\langle n_{\gamma} \rangle = \frac{1}{N} \sum_{i,\sigma} \langle n_{i\sigma\gamma} \rangle.
\end{equation}
We also calculate the spin-spin correlations by using the Fourier transform of the real space $\langle \mathbf{S}_{i} \cdot \mathbf{S}_{j} \rangle$,
\begin{equation}
S(k_{z},k_{y}) = \frac{1}{N^{2}} \sum_{i,j} e^{-i\vec{k}\cdot\vec{r}_{ij}} \langle \mathbf{S}_{i} \cdot \mathbf{S}_{j} \rangle,
\end{equation}
where $\textbf{S}_{i} = \sum_{\gamma} \textbf{S}_{i\gamma}$ (sum over the orbitals). 
Below, this spin structure factor will carry a subindex ``L'' or ``S'' depending on whether
in the Hamiltonian the long-range or short-range hoppings are used, respectively.

To explore pairing tendencies, we study the binding energy of a pair of holes defined as \cite{RMP94}
\begin{equation} \label{eq:bindingEnergy}
\Delta E = E(N-2) + E(N) - 2 E(N-1),
\end{equation}
where $E(M)$ is the ground state energy of the model with a total of $M$ electrons ($M=N$ is half-filling). If 
the particles minimize their energy by creating a bound state then $\Delta E$ is negative; if the holes become two independent particles this corresponds to zero binding energy in the bulk limit. In the case where the particles do not bind, this quantity is positive for finite systems.

To study the effects of holes on the magnetic correlations, we define a projector $P_{h\gamma}(i)$ 
at site $i$ such that it projects out the portion of the ground-state in which site $i$ and orbital $\gamma$ is occupied~\cite{S.White_holepair_TJ}: 
\begin{equation}\label{eq:HoleProjector}
 P_{h\gamma}(i) = c_{i\downarrow\gamma} c^{\dagger}_{i\downarrow\gamma} c_{i\uparrow\gamma} c^{\dagger}_{i\uparrow\gamma}.
\end{equation}
In order to work in the Hilbert space corresponding to $N_h$ number of holes 
at specific locations, we apply a product of projectors onto the ground 
state with $N_h$ holes, $P_{h\gamma} = P_{h\gamma}(i_1)P_{h\gamma}(i_2)...P_{h\gamma}(i_{N_h})$, where $i$ is the site to be projected while respecting the fermionic normal ordering ($i_i < i_2 < ... < i_{N_h}$). For example, $P_{ha} = P_{ha}(6)P_{ha}(8)$ projects out the occupied part of the ground state 
on orbital $a$ at sites $6$ and $8$. In fact, for most results shown below, 
we only apply the projector onto orbital $a$ in order to observe the corresponding local spin-spin correlations $\langle \psi | {\mathbf{S}_{ia} \cdot \mathbf{S}_{ja}} P_{ha}| \psi \rangle / \langle \psi | P_{ha}| \psi \rangle$, where the maximum possible magnitude of the correlations is $3/4$.

\begin{figure}[!ht]
\centering{
\includegraphics[width=8.0cm, height=14.0cm, clip=true]
				{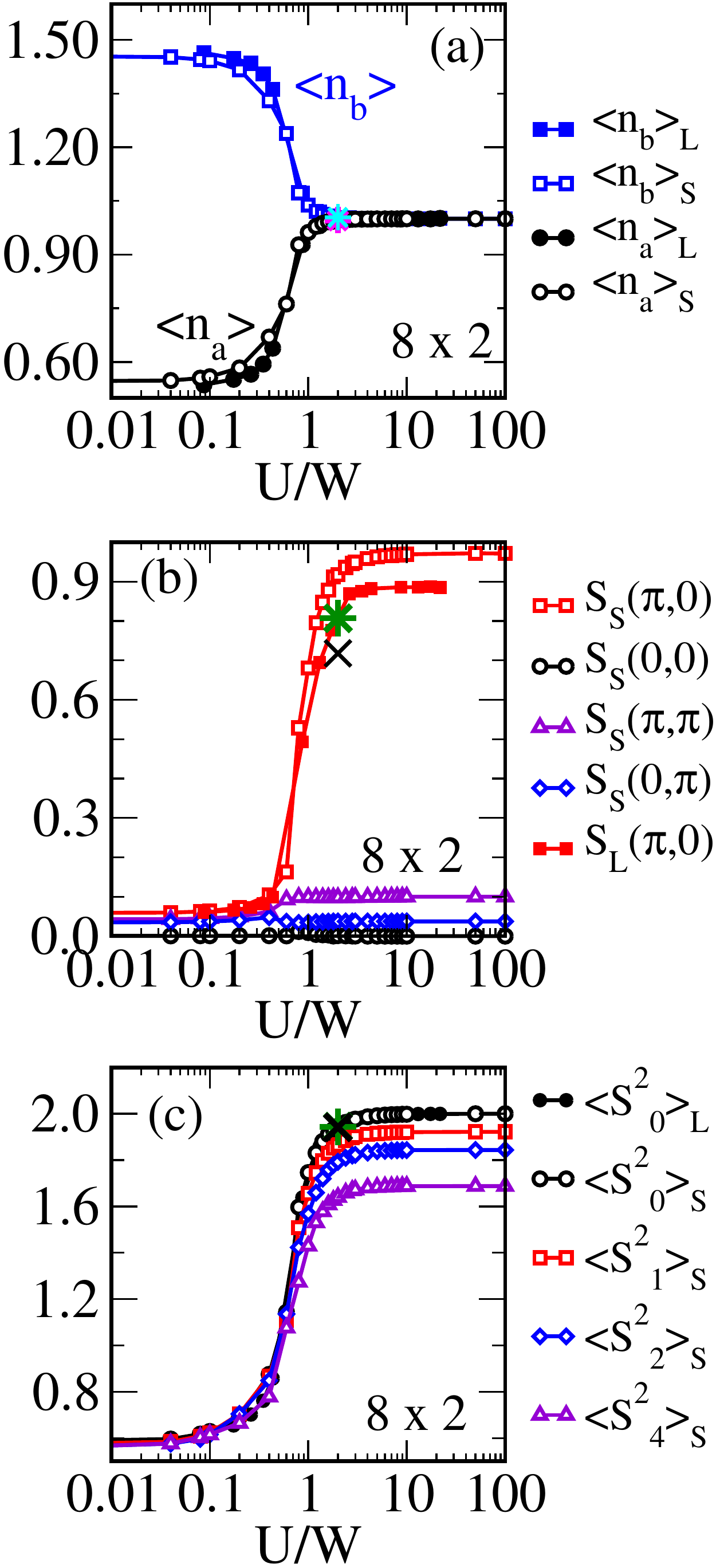}
		  }
\caption{(color online) Charge and magnetic properties of a half-filled 
$8 \times 2$ ladder at $P = 12.36$ GPa (Eq.~\ref{Eq3:P12}) studied with DMRG. 
Full (empty) points with subscript ``L'' (``S'') correspond to 
using long-range (short-range) hoppings. All results are at $J/U=0.25$. 
(a) Average orbital occupation vs. $U/W$. Black (blue) color 
is for orbital $a$ ($b$). 
The stars at $U/W = 2.0$ indicate results using a 
$12 \times 2$ lattice [cyan (magenta) for orbital $b$ ($a$)]. Very
similar results were obtained for a $16\times2$ lattice (not shown).  
(b) Fourier transform of the spin-spin correlations (i.e. spin structure factor) 
at representative wavevectors, indicating the dominance
of $(\pi,0)$ (see text).
At $U/W = 2.0$, $S_S(\pi,0)$ was also calculated using 
$12 \times 2$ (green star) and $16 \times 2$ (black X) lattices. 
$S_S(\pi,0)$ slightly decreases with system size because the 
one dimensional nature of the lattice prevents long-range magnetic order.
(c) $\langle S^2 \rangle$ vs. $U/W$, averaged over all sites. 
The subindexes 0, 1, 2, and 4 are the number of holes away from half-filling.
The green star and black X are as in (b), suggesting small size effects.
The convergence to 2 with increasing $U/W$ at half-filling denotes a convergence
to spin $S=1$, as expected because $J$ increases proportional to $U$. 
$\langle S^2 \rangle$ slightly decreases with increasing number of holes because of
dilution effects.
}
\label{Fig4}
\end{figure}

\section{Results}

This section presents our main results. We start with
the half-filled case that should be contrasted with the experimental data for the two-leg 
BaFe$_2$S$_3$  at pressures where magnetic order was reported. 
\emph{The magnetic order observed experimentally emerges very clearly from
our calculations.} We then proceed to the addition
of holes, under the assumption that the high pressure used experimentally 
moves bands around in such a manner that
the two-leg ladders become effectively doped. Our main result is that indications
of pairing are found in small systems, opening the possibility that indeed 
superconducting tendencies may be present in the models studied here.


\subsection{Half-Filling}

Figure~\ref{Fig4}~(a) shows
the electronic population of the two orbitals calculated via DMRG 
as a function of $U/W$, for the two electrons
per site half filling case. Because of the crystal
field splitting that locates orbital $b$ approximately 0.7~eV below
orbital $a$, in the weak coupling regime orbital $b$ is considerably
more populated. As the energy penalization for double occupancy
increases with increasing $U/W$, eventually at $U/W \sim 1$ both orbitals become
effectively singly occupied. These orbital populations are robust varying
the lattice size and also using either the ``short'' or ``long'' version 
of the hopping amplitudes, as shown in Fig.~\ref{Fig4}~(a).

Figure~\ref{Fig4}~(b) presents
the spin structure factor at various wavevectors
as a function of $U/W$. 
The wavector $(\pi,0)$ clearly dominates, particularly in the regime of
intermediate and strong coupling. 
$S_S(\pi,0)$ (see definition in caption of Fig.~4) starts growing already at $U/W \simeq 0.4$ even before
the full moments are developed, an intermediate
coupling regime that several investigations assign to the
iron based superconductors~\cite{Dainature}. Once again, these results are robust
increasing the lattice size and using either ``short'' or ``long'' hopping
amplitudes. The small decrease of $S_S(\pi,0)$ in Fig.~\ref{Fig4}~(b)
with increasing clusters from 8$\times$2 to 16$\times$2 is reasonable
because a true long-range order is not expected in one dimension, but
a slow power-law decay should instead prevail. 

The dominance of the $(\pi,0)$ magnetic order is in excellent
agreement with neutron experiments for BaFe$_2$S$_3$~\cite{NatMatSC}. In our
model, this magnetic order dominance arises primarily from the
comparable strength of the hopping amplitudes $t^{y+z}$
along the diagonal of the elementary plaquettes contrasted with those along the
nearest-neighbor sites along the rungs and legs. This comparable
strength originates in the location of sulphur, that acts 
as a bridge between irons, up and down the middle of 
the ladder plaquettes. This is also the same reason 
for the dominance of the (degenerate) $(\pi,0)$ and $(0,\pi)$ 
wavevectors in planar geometries, at intermediate 
and strong couplings.

In two-leg ladders the explicit breaking of the lattice
rotational invariance renders $(\pi,0)$ and $(0,\pi)$ 
no longer degenerate. But why $(\pi,0)$ dominates over 
$(0,\pi)$ according to the DMRG calculations? A possible
simple explanation is the following. Consider a classical
$J_1$-$J_2$ spin model for spins of magnitude 1, where 
$J_1$ is the antiferromagnetic Heisenberg coupling for
nearest-neighbors spins both along the rungs and legs,
while $J_2$ is the antiferromagnetic coupling along the
plaquette diagonals. The
energy of the $(\pi,0)$ state is always smaller than the
energy of the $(0,\pi)$ because in $(\pi,0)$
each spin always has two nearest-neighbors AFM links, while in $(0,\pi)$ there is
only one nearest-neighbor AFM link. While at small $J_2/J_1$ the $(\pi,\pi)$ order
dominates as expected,  a level crossing to $(\pi,0)$
eventually occurs at $J_2/J_1 = 0.5$. This also provides a possible
rationale for why $(\pi,\pi)$ rather than $(0,\pi)$
appears to be the subdominant order in Fig.~\ref{Fig4}~(b): in these two-orbital
Hubbard models for two-leg ladder materials
the ratio $J_2/J_1$ between the effective Heisenberg couplings in strong coupling
must be between 0.5 and 1.0. 

Figure~\ref{Fig4}~(c) plots the spin squared expectation
value as a function of $U/W$, showing the formation of local moments. The
upturn with increasing $U/W$ occurs at values similar to those
where $S(\pi,0)$ starts growing.
Eventually at strong coupling, $U/W > 1$, the spins
are fully developed and they acquire their maximum value $S=1$ i.e.
a magnetic moment 2.0 $\mu_B$.
Neutron scattering experiments at ambient pressure~\cite{NatMatSC}
 report a moment of 1.2 $\mu_B$ ($S \sim 0.6$),
which we find at $U/W \simeq 0.5$; first principles 
predict a value of 2.0 $\mu_B$ ($S \sim 1.0$) 
at the same pressure~\cite{SuzukiPRB15}.
Note that neutron scattering may be capturing a moment that is time
averaged, thus reducing its value, and other techniques should be used
to find the actual instantaneous spin~\cite{moments1,moments2}.
Also note that comparing magnetic moment results of a two orbital model vs. calculations
and experiments involving five orbitals is difficult.
Regardless, intermediate to strong coupling
is the physically relevant regime in this model from the magnetic moment
perspective.

\begin{figure}[!ht]
\centering{
\includegraphics[width=8.0cm, height=10.0cm, clip=true]
				{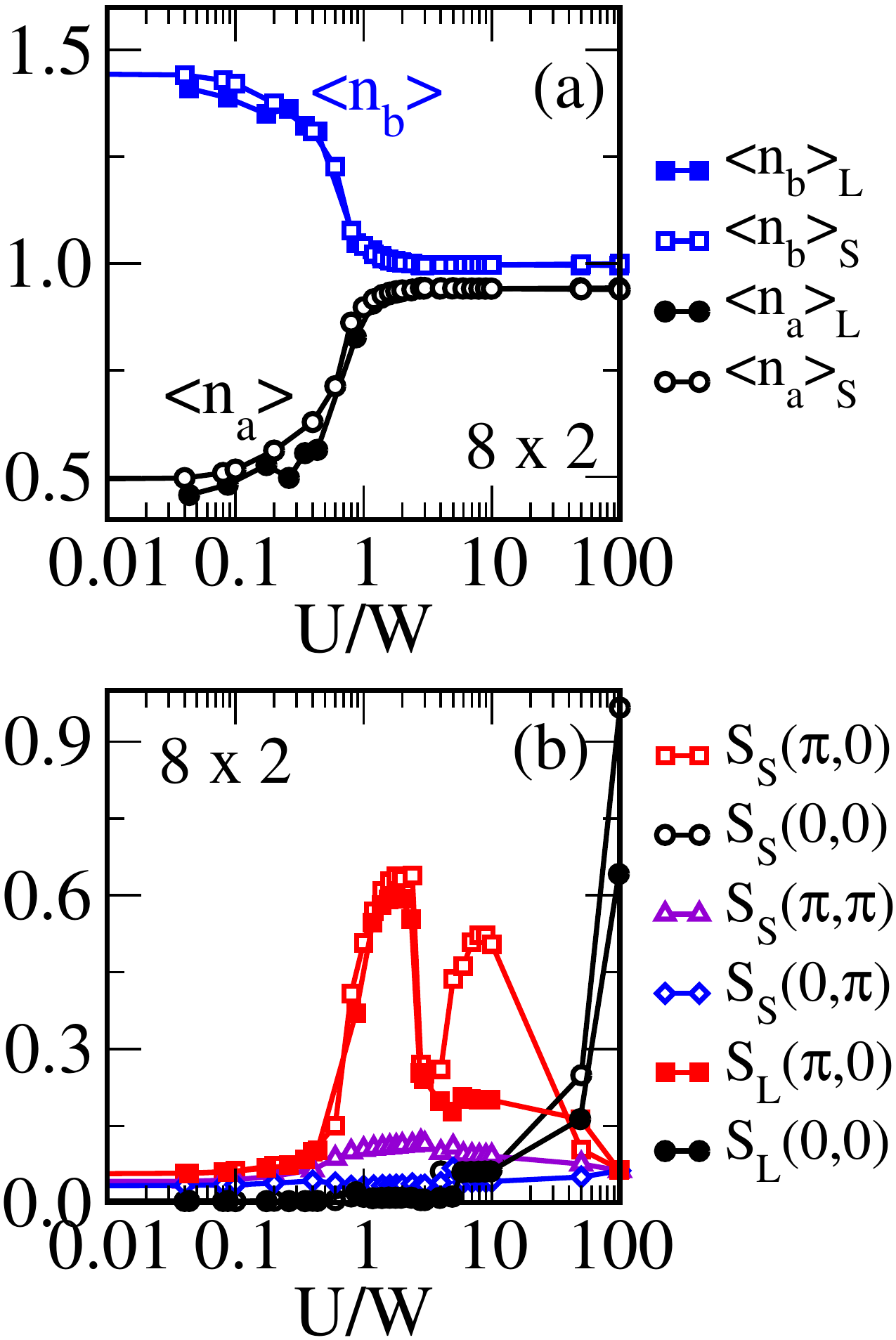}
		  }
\caption{(color online) Charge and magnetic properties of 
an $8 \times 2$ ladder doped with one hole, 
at $P = 12.36$ GPa (Eq.~\ref{Eq3:P12}) and studied with DMRG. 
Full (empty) points with subscript ``L'' (``S'') correspond to 
using long-range (short-range) hoppings. All results are at $J/U=0.25$. 
(a) Average orbital occupancy vs. $U/W$. Black (blue) color 
is for orbital $a$ ($b$). At small $U/W$ the crystal
field creates a substantial difference in the populations. At large
$U/W$ the $b$ orbital converges approximately to one electron/site, 
while the $a$ orbital contains most of the doped hole. The results 
are approximately the same for L and S hoppings.
(b) Fourier transform of the spin-spin correlations (i.e. spin structure factor) 
at various representative wavevectors, indicating the dominance
of $(\pi,0)$. At very large $U/W$ the one hole state becomes ferromagnetic
due to double exchange tendencies, as discussed in the text.
}
\label{Fig5}
\end{figure}


\subsection{One Hole Doped}

Figure~\ref{Fig5} shows
results for the case of one hole doped into the half-filled system. 
Panel (a) displays the population
of each orbital. As at half-filling, the crystal field splitting
induces a large difference at weak coupling
between the two orbitals.
However, it is curious to observe that in the strong
coupling regime the hole is still almost entirely located at orbital $a$,
in spite of the presence of a gap induced by the repulsion $U$. Nevertheless,
since the $U$ is the same for both orbitals, the only asymmetry between the
orbitals is the original crystal field splitting that, therefore, 
must be inducing the asymmetric population with holes of the $a$ orbital.

This strong-coupling Wannier orbital population, where one orbital is locked
at one electron/site and the other at less than one electron/site,
corresponds to an orbital selective Mott phase (OSMP)~\cite{OSMP}. In this
context orbital $b$ provides localized spins $S=1/2$, that
are in interaction with delocalized carriers at orbital $a$. The physics
of the OSMP state suggests that this state, if realized in the present
two-leg ladders, may have exotic transport properties that include
a very small quasiparticle weight.

\begin{figure}[!ht]
\centering{
\subfloat{
\includegraphics[width=8.0cm, height=2.5cm, clip=true]
				{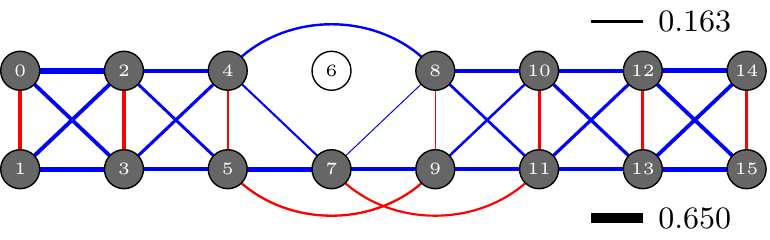}
	\llap{\parbox[b]{15cm}{\large\bf{(a) $8 \times 2$}\\\rule{0ex}{2.0cm}}}
}\\
\subfloat{
\includegraphics[width=8.0cm, height=2.5cm, clip=true]
				{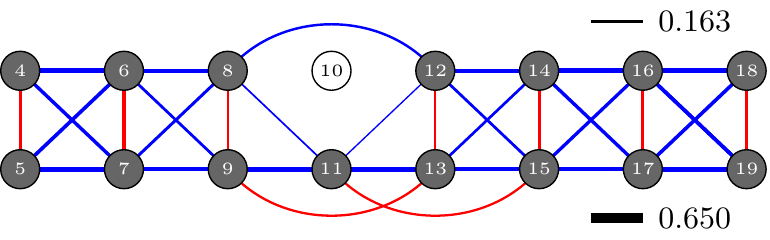}
	\llap{\parbox[b]{15cm}{\large\bf{(b) $12 \times 2$}\\\rule{0ex}{2.0cm}}}
}
}
\caption{(color online) Results obtained from the wave function of a 
dynamical hole at $P = 12.36$ GPa (Eq.~\ref{Eq3:P12}), using $U/W=2$ and $J/U=0.25$, 
for the case when the hole is projected on orbital $a$ 
at the location denoted by the white circles (see Eq.~\ref{eq:HoleProjector}).
(a) are results for the $8\times2$ lattice and (b) are for the $12\times2$ lattice,
showing that size effects are small.
The thickness of the lines is linearly proportional to the magnitude of the spin-spin 
correlations involving orbital $a$. These correlations between spins at sites $m$ and $j$
are defined as $\langle \psi| \mathbf{S}_{ma} \cdot \mathbf{S}_{ja} P_{ha}(i)| \psi \rangle / \langle \psi| P_{ha}(i)| \psi \rangle$, 
where $P_{ha}(i)$ was defined in the text.
Blue denotes antiferromagnetic correlation while red is ferromagnetic. 
In (a) $i = 6$, and in (b) $i = 10$. Magnetic correlations away from the hole are very 
similar to those in the undoped case. 
In both panels (a) and (b) in the vicinity of the hole a weak antiferromagnetic correlation 
between spins ``across the hole'' location can be clearly observed. For a discussion see the text.
}
\label{Fig6}
\end{figure}

Panel (b) contains the spin structure factor. As in the case
of half-filling, clearly the wavevector $(\pi,0)$ dominates starting
at $U/W \sim 0.4$, and irrespective of using ``short'' or ``long''
range hoppings. The dip at $U/W \sim 3$ is unexpected and it may
reflect on how the hole scrambles the original magnetic order
as the size of the spin distortion around the hole changes with $U/W$.
This spin scrambling effect can be better visualized in Fig.~\ref{Fig6}
where from the entire wave function of the one-hole state, a projection
is made for the case where the hole is located at the sites indicated.
While far from the projected hole the spin order is basically unchanged
from the half-filled $(\pi,0)$ pattern, in the vicinity of the hole
there is an inevitable scrambling effect that broadens the $(\pi,0)$ peak. 
This shift of weight away from $\pi$ along the leg direction 
is exemplified by the antiferromagnetic coupling ``across the hole'' involving
e.g. spins 4 and 8 on the 8$\times$2 lattice that otherwise should be
ferromagnetically coupled. The 12$\times$2 results in the same panel
indicate very small size effects. This across-the-hole AFM coupling
has been observed in the $t-J$ model context before~\cite{martins1,martins2,martins3}, and it is considered
a precursor of spin-charge separation at least at short distances.
In fact, the exact ground state of the $U=\infty$ one-orbital Hubbard model
in one dimension presents an exact decoupling between spin and charge with
AFM couplings across all holes~\cite{shiba}.

We warn the readers that there are some qualitative differences
between the cases of short and long range hoppings. Of instance in Fig.~\ref{Fig5}~(b)
$S_S(\pi,0)$ has a ``second peak'' at $U/W \sim 10$ which is suppressed in
$S_L(\pi,0)$. We do not know the qualitative reasons for this difference. However,
in the important region of pair binding $U/W \sim 2$, 
to be described later in the text, both short and long range hoppings give very similar
results.

An interesting observation from Fig.~\ref{Fig5}~(b) is that
at very large $U/W$ eventually the one hole state becomes
ferromagnetic since $S(0,0)$ dominates. In multiorbital systems, especially in cases
where some degrees of freedom are localized and others itinerant
as it occurs in this model, double exchange mechanisms can favor ferromagnetic
tendencies as it occurs in manganites~\cite{manganites}. In the large $U/W$ regime,
the effective Heisenberg couplings $J_1$ and $J_2$ are very small
since they are inversely proportional to $U$,
while the Hund coupling being fixed to $J/U=0.25$ is very large.
Such a regime is clearly favorable for double exchange tendencies, as
shown by the DMRG results. This also indicates that ferromagnetic states
are close in parameter space to the realistic regimes for iron superconductors,
a conclusion that also emerged from previous investigations~\cite{Feladder7,luo2D,luo1D}.

\begin{figure}[!ht]
\centering{
\includegraphics[width=8.0cm, height=10.0cm, clip=true]	
				{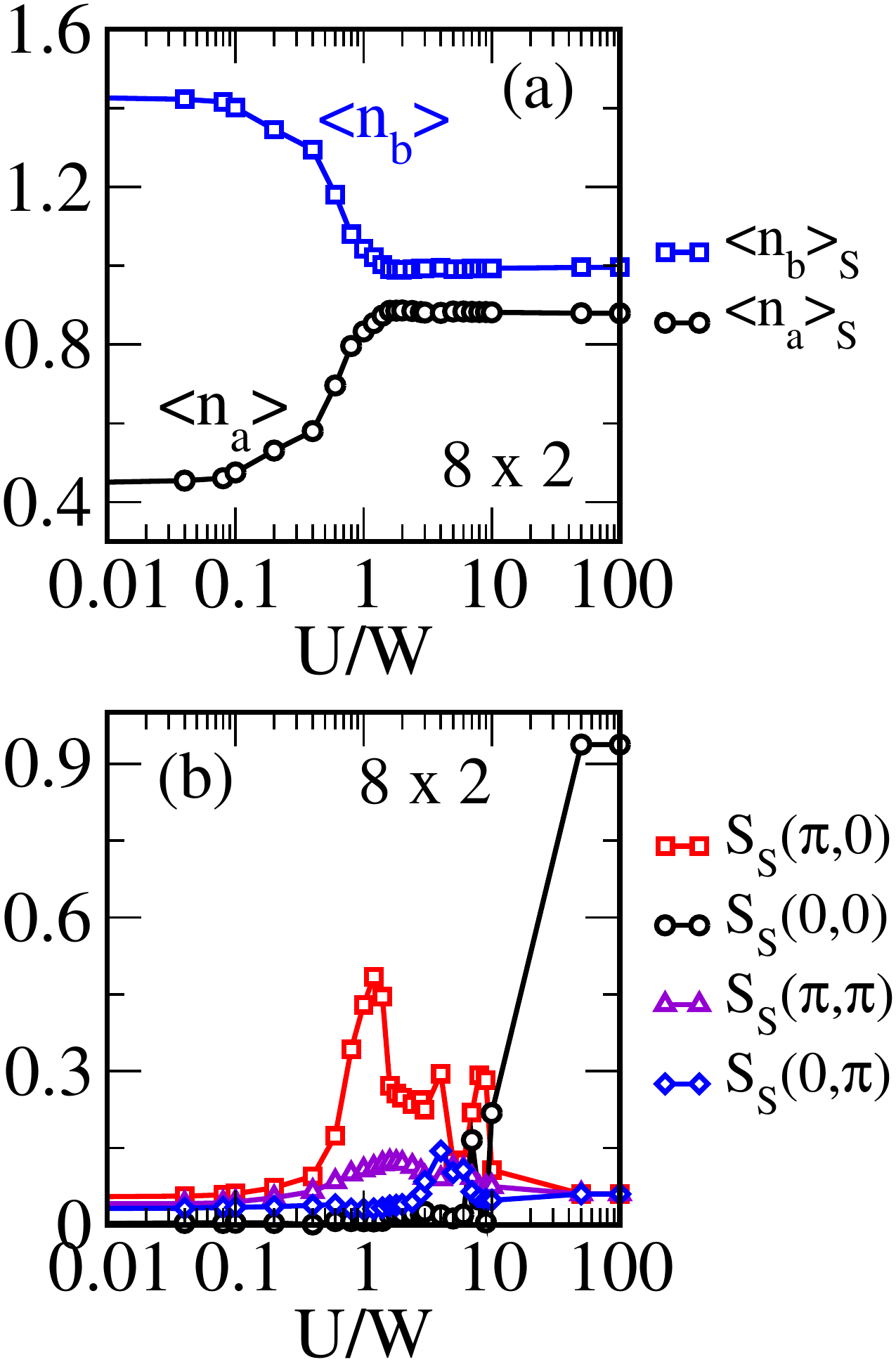}
		  }
\caption{(color online) 
Charge and magnetic properties of 
an $8 \times 2$ ladder doped with two holes, 
at $P = 12.36$ GPa (Eq.~\ref{Eq3:P12}), using short range (``S'') hoppings, and studied with DMRG. 
All results are at $J/U=0.25$. 
(a) Average orbital occupancy vs. $U/W$. Black color 
is for orbital $a$ and blue for orbital $b$. At small $U/W$ the crystal
field creates a substantial difference in the populations. At large
$U/W$, the $b$ orbital converges approximately to one electron/site, 
while the $a$ orbital contains most of the doped holes.
(b) Fourier transform of the spin-spin correlations (i.e. spin structure factor) 
at representative wavevectors, indicating the dominance
of $(\pi,0)$ at intermediate/large $U/W$. 
At very large $U/W > 10$ the two holes state becomes ferromagnetic
due to double exchange tendencies, as discussed in the text and as for one hole.
}
\label{Fig7}
\end{figure}

\subsection{Two Holes Doped}

The results for two doped holes shown in Fig.~\ref{Fig7} continue
the trends observed before for one hole. Panel (a) shows the Wannier
orbital populations as a function of $U/W$. As for one hole, at 
large $U/W$ the orbital $b$ population remains locked at one electron/site,
while the two holes almost entirely reside at orbital $a$. This confirms
the tendency towards an OSMP state with doping. With regards to the
spin magnetic order, panel (b), the $(\pi,0)$ order still dominates
in the broad region between $U/W=0.4$ and $10$, but the spin order
scrambling caused by the mobile holes reduces the intensity of
$S(\pi,0)$ as expected. In addition, the tendency towards ferromagnetism
triggered by double exchange continues at very large $U/W$.


\begin{figure}[!ht]
\centering{
\includegraphics[width=8.0cm, height=10.0cm, clip=true]
				{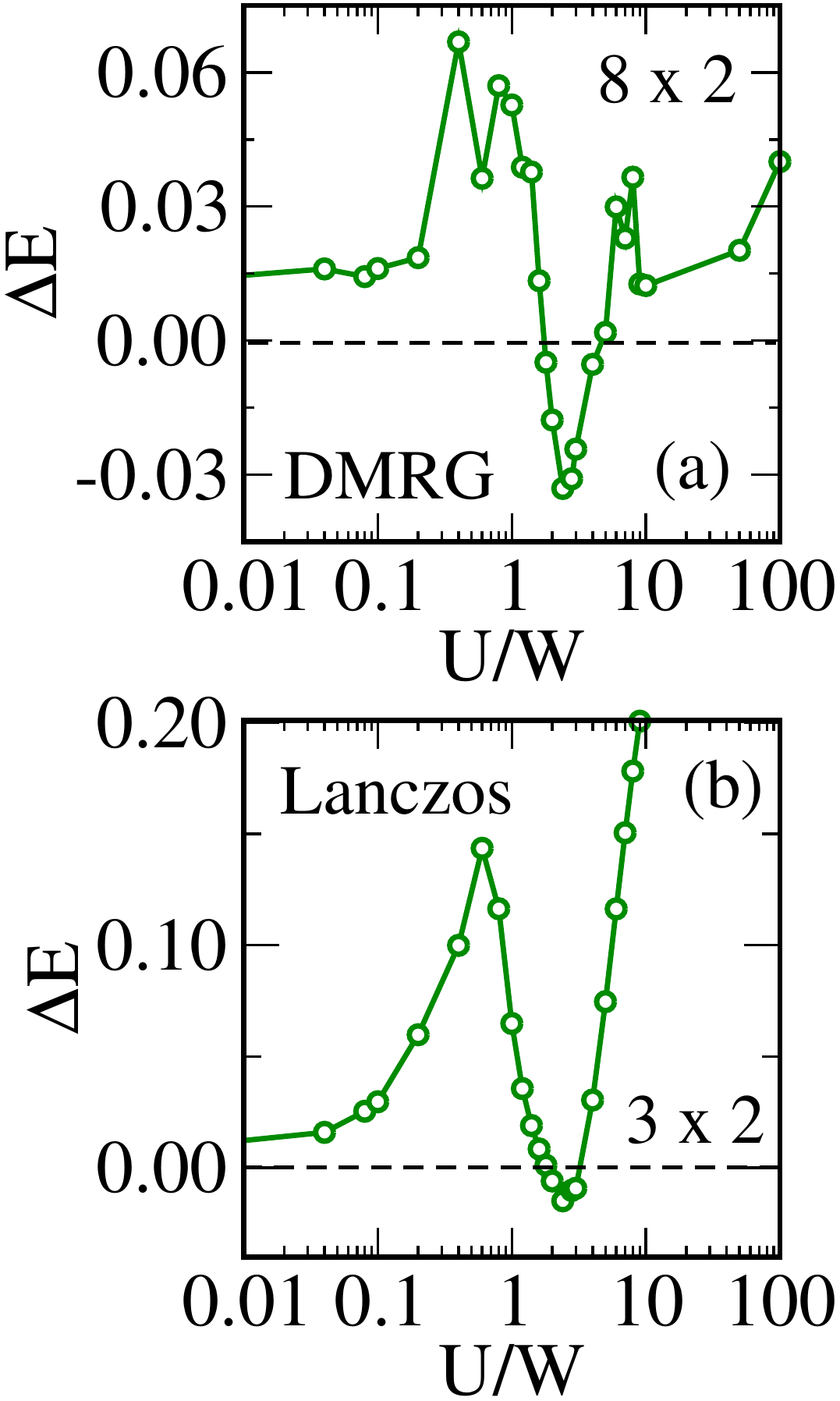}
		  }
\caption{(color online) Binding energy vs. $U/W$ calculated 
using (a) DMRG for an $8\times2$ ladder and (b) Lanczos for 
a $3\times2$ ladder. 
In both cases, we observe a non-monotonic up-down-up behavior where the 
minimum of the binding energy can be found at $U/W \sim 2$. Since 
this minimum of $\Delta E$ is negative, panel (a) suggests 
binding of holes between $U/W \sim 1.5$ 
and $U/W \sim 4.5$. The results in both panels 
were obtained at $P = 12.36$ GPa (Eq.~\ref{Eq3:P12}), using short-range 
hoppings, and $J/U=0.25$. 
}
\label{Fig8}
\end{figure}

\subsection{Binding Energy}

After calculating the ground state energies for the $N$, $N-1$ (1 hole),
and $N-2$ (2 holes) subspaces, we can also calculate the binding
energy $\Delta E$ previously defined. The remarkable result shown in
Fig.~\ref{Fig8}~(a) is that
for the 8$\times$2 cluster this quantity becomes negative between
$U/W \sim 1.5$ and $U/W \sim 4.5$. This is a broad region, in spite
of the perceived narrowness in panel (a) because of the logarithmic scale
used. In this regime the spins are already well developed and near
saturation as it was shown in Fig.~\ref{Fig4}~(c).
Considering that holes are
located  at orbital $a$, this surprising result brings similarities with negative
binding energies found in one-orbital models for the cuprates, 
such as the $t-J$~\cite{RMP94}. In fact, the crude rationale for binding based
on the ``number of broken AFM links'' may apply here as well~\cite{RMP94}. In this context,
binding occurs because each hole damages the AFM spin state,
and the manner to minimize the size of that distorted magnetic background is
by bringing the holes together. 
It is also interesting that exact results obtained via the Lanczos method applied
to a very small 3$\times$2 lattice produce a profile for the binding
energy, shown in Fig.~\ref{Fig8}~(b), that qualitatively resembles 
panel (a) suggesting that size effects are mild. Alas, as already 
explained, we have not been able to reach sufficient accuracy in the two holes
sector to confirm the pairing tendencies of Fig.~\ref{Fig8}~(a)  with larger
lattices, thus our pairing analysis below is restricted to the 8$\times$2 cluster.

\begin{figure}[!ht]
\centering{
\includegraphics[width=8.7cm, height=10.0cm, clip=true]
				{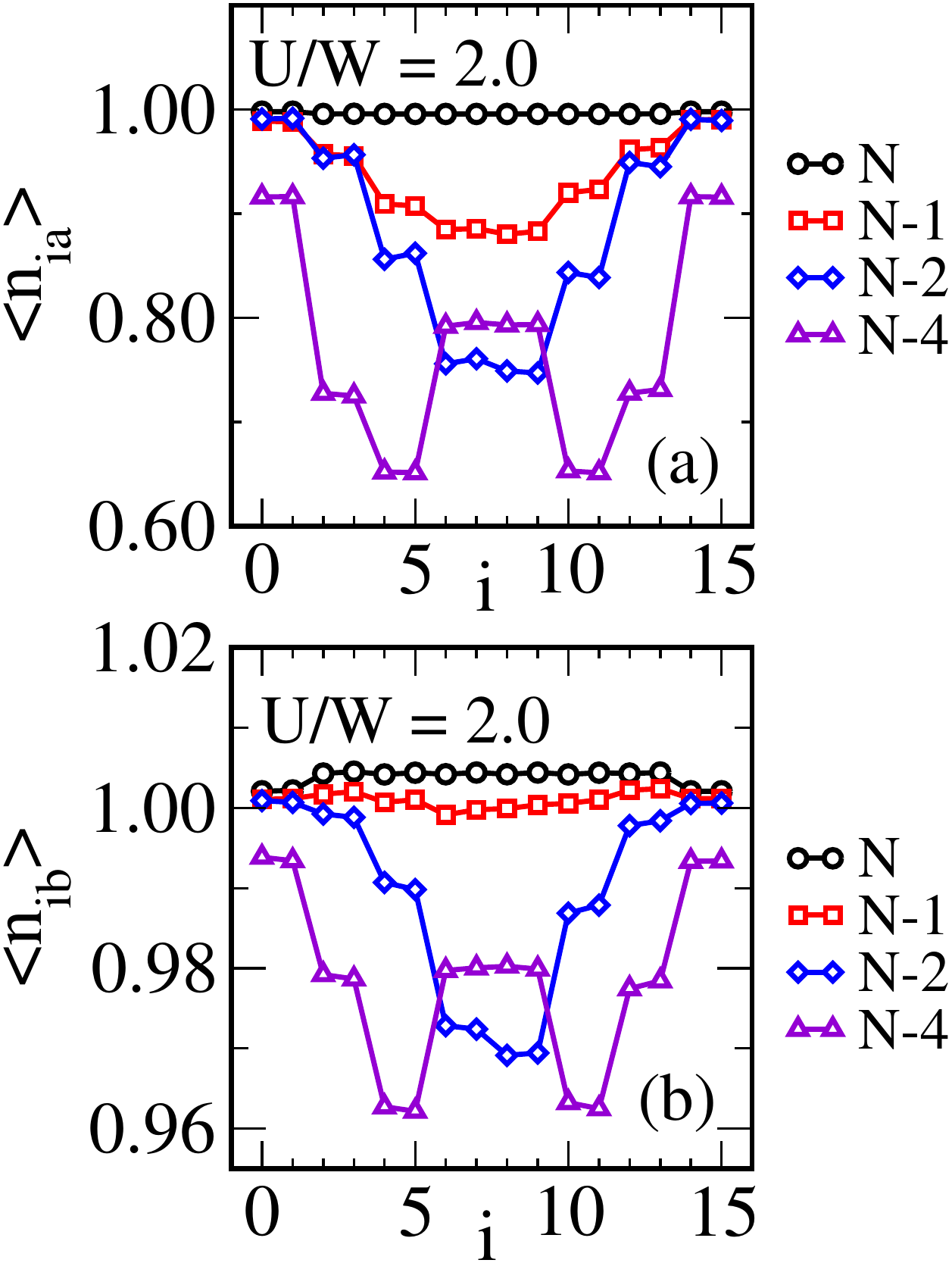}
		  }
\vspace{-0.4cm}
\caption{(color online) Real space electronic density of each orbital 
[panel (a) is for orbital $a$, while panel (b) for orbital $b$] 
using an $8\times2$ ladder, $U/W = 2.0$, $J/U=0.25$, 
short-range hoppings, and at $P = 12.36$ GPa (Eq.~\ref{Eq3:P12}). 
Results are shown for
half-filling ($N$ electrons), one hole ($N-1$), two holes ($N-2$), and four holes ($N-4$) 
as a function of the position ``$i$'' (see Fig.~\ref{Fig6}~(a) for 
the site labeling convention). The most stricking result corresponds
to four holes where the presence of two minima is indicative of hole pairing.
Doping of four holes reduces the orbital $a$ electron density
by approximately $25\%$, while orbital $b$ has a charge depletion of 
only $ \sim 3\%$, illustrating again that holes mainly reside at orbital $a$.
}
\label{Fig9}
\end{figure}

The results in panel (a) suggesting pairing in a region of parameter space 
brings analogies with the
negative binding energies reported before in 
Kondo lattice models for heavy fermions~\cite{xavierPRL}. For instance
in Fig.~2~(b) of Ref.~\cite{xavierPRL} a negative $\Delta E$ 
is reported using up to 32$\times$2 lattices. Even the pair-pair
correlation functions of Fig.~3~(b) of Ref.~\cite{xavierPRL} 
(unfortunately not within the reach
of the present study that uses the full two-orbital Hubbard model) 
suggest a dominant pairing tendency in the doped Kondo lattice on two-leg ladders.
Perhaps having holes primarily at orbital $a$ (as shown before), while orbital $b$ remains singly occupied
at strong coupling, effectively transforms our model into a Kondo lattice model.

\begin{figure}[!ht]
\centering{
\subfloat{
\includegraphics[width=8.2cm, height=4.5cm, clip=true]
				{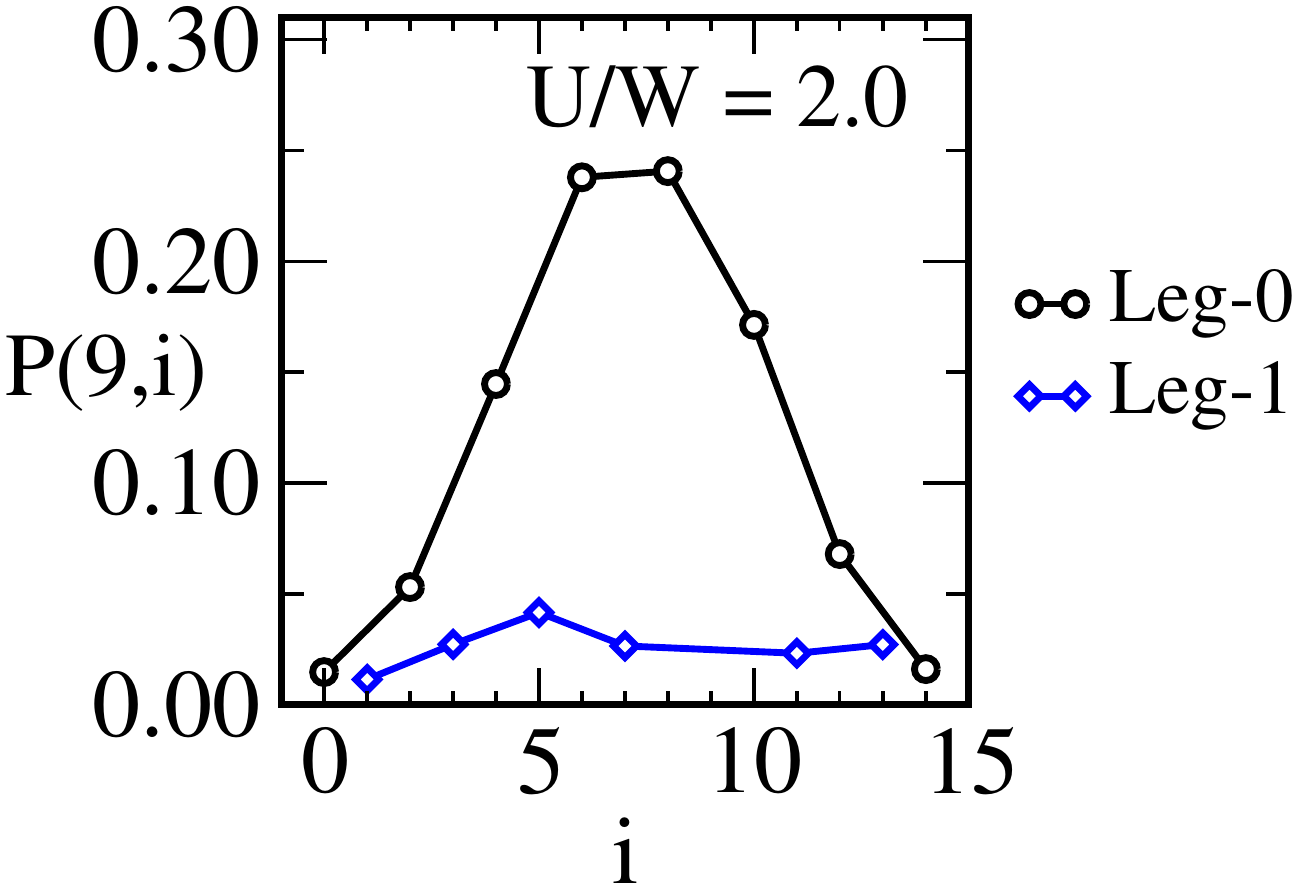}
	\llap{\parbox[b]{11.3cm}{\Large\bf{(a)}\\\rule{0ex}{3.5cm}}}
}\\
\vspace{-0.4cm}
\subfloat{
\includegraphics[width=8.0cm, height=2.5cm, clip=true]
				{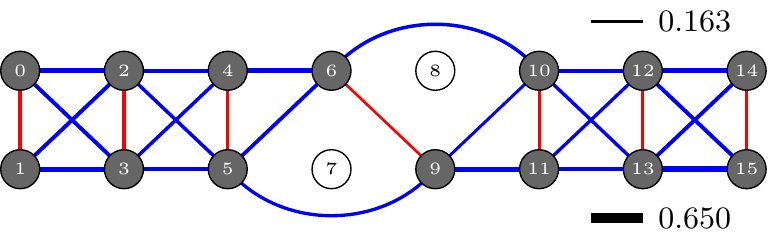}
	\llap{\parbox[b]{15cm}{\Large\bf{(b)}\\\rule{0ex}{2.0cm}}}
}\\
\vspace{-0.2cm}
\subfloat{
\includegraphics[width=8.0cm, height=2.5cm, clip=true]
				{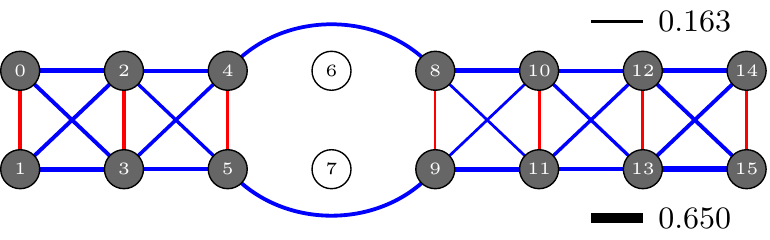}
	\llap{\parbox[b]{15cm}{\Large\bf{(c)}\\\rule{0ex}{2.0cm}}}
}
}
\caption{(color online) 
(a) For the case of the two holes ground state on an $8\times2$ 
ladder, this panel shows the probability of a hole to be located at 
a site ``$i$'' assuming the other hole is fixed at site $9$ of 
the bottom leg (Leg-1). Results are normalized to one. Since sites are labeled with a snake-like 
geometry where site zero starts from the upper leg (Leg-0), this panel indicates
that the two holes in the bound state are primarily located in different legs.
See Fig.~\ref{Fig6}~(a) for site labelling details.
(b,c) Results from the DMRG ground state 
wave function of two holes on an 8$\times$2 cluster, using 
$U/W = 2.0$ (binding region) 
and $J/U=0.25$, and with short-range hoppings at $P = 12.36$ GPa (Eq.~\ref{Eq3:P12}).  
Shown are spin-spin correlations for the case where the two holes are projected to be
at the white circles, i.e. (b) along a plaquette diagonal and (c) along a  rung. These
are the two most dominant configurations in the hole pair.
The spin-spin magnetic correlations are defined as 
$\langle \psi| \mathbf{S}_{ma} \cdot \mathbf{S}_{na} P_{ha}(i) P_{ha}(j)| \psi \rangle / \langle \psi| P_{ha}(i) P_{ha}(j)| \psi \rangle$, involving only orbital $a$ because it is the primary
location for the doped holes. 
In (b) $i = 7$ and $j = 8$ and in (c) $i = 6$ and $j = 7$.
Blue (red) lines are AFM (FM) bonds. 
For all hole configurations, the $(\pi,0)$ magnetic order 
is substantially distorted only near the holes. 
Note also the presence of ``across the hole'' AFM correlations in both panels.
}
\label{Fig10}
\end{figure}

To further test the pairing implication of finding a negative binding
energy in Fig.~\ref{Fig8}, we have also analyzed the real space distribution of
holes in the doped system. In Fig.~\ref{Fig9}~(a), the electronic density
is shown for orbital $a$, where the holes are mostly located, 
for each of the 16 sites of the 8$\times$2 lattice
at coupling $U/W=2$ where $\Delta E$ is negative. For $N$ electrons, i.e.
half-filled, the electronic density is basically uniform. In the case of
$N-1$ electrons, i.e. one hole, this hole is located in the middle of the
cluster as expected for a system with open boundary conditions. For the
case of two holes corresponding to $N-2$ electrons, these two holes are 
also located near the center of the cluster but in a tight manner compatible with
pairing. The most important result is for the case of four holes, corresponding to
$N-4$ electrons, since Fig.~\ref{Fig9}~(a) indicates the presence of
$two$ minima in the electronic density, a result compatible with the presence
of two hole pairs,  as opposed to a single broad minimum which would 
indicate independent holes or four minima which would signal 
a charge density wave of holes. There are also no indications of phase separation. 
The results in Fig.~\ref{Fig9}~(b) for orbital $b$
simply mirror those of orbital $a$ but with a far more suppressed hole density.

Figure~\ref{Fig10}~(a) illustrates the internal structure of the hole pairs
that we have found using the 8$\times$2 cluster. This figure is based on the wave
function of two holes, with one of the holes projected to site ``9'' which is
on ``Leg-1''. This panel shows that the second hole is primarily located on
the other leg, i.e. ``Leg-0'', mainly at the sites either in the same rung as ``9'' or
diagonally across the plaquettes. Projecting now the two holes to those particular
locations, these two dominant ``plaquette diagonal'' and ``rung'' states
for the pair of holes are shown in Figs.~\ref{Fig10}~(b) and (c), respectively.
Similarly as for the case of one hole, there is a notorious ``across the hole'' antiferromagnetic
coupling between spins that otherwise should be ferromagnetically coupled
in the undoped system. This AFM correlation facilitates the movement of the hole. 
Note also that if in panel (b) the hole located at ``8'' and the spin at ``6''
are interchanged, as it would happen via the action of electronic hopping and
asssuming that the AFM and FM bonds remain the same as if they were elastic bands, 
then panel (c) is obtained. In fact, this panel (c) has an AFM across-the-hole coupling 
between ``4'' and ``8'' and a FM coupling between ``8'' and ``9'' that was 
originally a FM coupling along the diagonal from ``6'' to ``9'' in panel (b). 
Then panels (b) and (c) are compatible with one another with regards to hole pairing:
the two holes are oscillating in different legs close to one another due to an
attraction created by the antiferromagnetic background.

\section{Conclusions}

In this publication, we have presented the first study of a realistic (derived from first principles) 
electronic model Hamiltonian 
for the two-leg ladder compound BaFe$_2$S$_3$ that was recently shown to become superconducting at
high pressure~\cite{NatMatSC,ohgushi}. The model has two orbitals and electronic hoppings beyond nearest neighbor iron sites,
rendering its study difficult even with the powerful DMRG method. For this reason our analysis has been
restricted to relatively small clusters. Nevertheless,  
we have been able to extract interesting information
from the model that is in good agreement with experiments. For example, the parent compound has magnetic
order involving ferromagnetic rungs that are coupled antiferromagnetically along the legs, as found in neutron
scattering experiments~\cite{NatMatSC}. In the strong coupling limit, this order emerges from the competition
between antiferromagnetic Heisenberg couplings along rungs and legs and along the diagonals of the plaquettes. 
With hole doping, we observed
that only one of the two Wannier orbitals used here becomes populated. This indicates a tendency towards effective
models involving a combination of itinerant and localized orbitals, as in the context of an orbital
selective Mott phase.
Even more exciting, we have found that at strong coupling and using
an 8$\times$2 cluster with two holes, there are indications of hole pair formation induced by antiferromagnetism.
While this result must be confirmed using larger systems and more DMRG states, a challenging task, it
suggests that this type of two-orbital models  
contains the essence of the mechanism for superconductivity in iron-based
two-leg ladders, a mechanism that could be similar to that in layered systems. 
As a consequence, our
present effort paves the way and motivates further studies in this context. 
We believe that the theoretical and experimental 
analysis of iron-based two-leg ladders may prove to be as interesting and illuminating as the early 
studies in copper oxide two-leg ladders were for cuprate physics, providing a novel
playground in the context of iron-based high-$T_c$ superconductivity.

\section{Acknowledgments}

N.P., A.M., and E.D. were supported by the National Science Foundation Grant No.
DMR-1404375. N. P. was also partially supported by the U.S. Department of Energy
(DOE), Office of Basic Energy Science (BES), Materials
Science and Engineering Division.
Part of this work was conducted at the Center for Nanophase Materials Sciences, sponsored
by the Scientific User Facilities Division (SUFD), BES, DOE, 
under contract with UT-Battelle. A.N. and G.A. acknowledge support by the Early
Career Research program, SUFD, BES, DOE. Computer time was provided in part by resources 
supported by the University of Tennessee and Oak Ridge National Laboratory Joint Institute 
for Computational Sciences ({\it http://www.jics.utk.edu}).

\end{document}